\documentclass[12pt]{iopart}
\expandafter\let\csname equation*\endcsname\relax
\expandafter\let\csname endequation*\endcsname\relax
\usepackage{amsmath}

\usepackage{iopams}    
\usepackage{graphicx}   
\usepackage{caption}

\usepackage[numbers,sort&compress]{natbib}
\usepackage{hyperref}
\usepackage[normalem]{ulem}
\hypersetup{
	colorlinks=true,
	citecolor=blue,
	linkcolor=blue,
	urlcolor=blue}

\begin{document}

\title[]{Thermal amplification and melting of phases in spin-orbit-coupled spin-1 Bose-Einstein condensates}
\author{Ritu}
\ead{ritu.22phz0002@iitrpr.ac.in}
\address{Department of Physics, Indian Institute of Technology Ropar, Rupnagar-140001, Punjab, India}
\author{Rajat}
\ead{rajat.19phz0009@iitrpr.ac.in}
\address{Department of Physics, Indian Institute of Technology Ropar, Rupnagar-140001, Punjab, India}
\author{Arko Roy}
\ead{arko@iitmandi.ac.in}
\address{School of Physical Sciences, Indian Institute of Technology Mandi, Mandi-175075, H.P., India}
\author{Sandeep Gautam}
\ead{sandeep@iitrpr.ac.in}
\address{Department of Physics, Indian Institute of Technology Ropar, Rupnagar-140001, Punjab, India}

\begin{abstract}
We implement Hartree-Fock-Bogoliubov theory with Popov approximation for a homogeneous Raman-induced 
spin-orbit-coupled spin-1 Bose-Einstein condensate and investigate the effects of finite temperature ($T$) on the
ground-state phase diagram. We calculate the roton gap as a function of Raman coupling ($\Omega$) or quadratic Zeeman 
field strength ($\epsilon$) to extract the critical points separating the supersolid stripe phase
from the plane wave or zero-momentum phase at finite temperatures. We present a few representative finite-temperature 
phase diagrams for the system in the $T-\Omega$ and $T-\epsilon$ planes. Our observations indicate that the supersolid stripe
phase melts at finite temperatures. We also discuss the contrasting roles of quantum and thermal fluctuations in shifting
the phase boundary separating the supersolid stripe from the plane-wave phase. 
\end{abstract}


\section{Introduction}
In recent years, synthetic spin-orbit (SO) coupling has become a powerful tool for exploring
many-body physics, arising from the interplay between the modified single-particle spectrum
and interactions~\cite{galitski2013spin, Goldman_2014}. In ultracold quantum gases, 
SO coupling refers to the interaction between the quantum mechanical spin
of a particle and its spatial motion. 
An equal strength mixture of Rashba and Dresselhaus couplings has been experimentally realized by dressing hyperfine spin states of an atom using Raman lasers ~\cite{sorealize, NatCommun} or by using a modulating gradient magnetic field \cite{synthesis_so}. 
The one stand-out ground state phase that arises in an SO-coupled Bose-Einstein condensate (BEC) is the supersolid stripe (ST) phase.
Supersolidity is a fascinating phenomenon that combines superfluidity with crystallinity, resulting from the breaking of two continuous symmetries: U(1) gauge and
translational invariance~\cite{recati2023supersolidity}. Experimental observation of supersolidity has been
reported in various systems, including SO-coupled BECs ~\cite{li2017stripe, PhysRevLett.124.053605, chisholm2024probing, 10.21468/SciPostPhys.11.5.092}, dipolar quantum gases~\cite{PhysRevLett.122.130405, PhysRevX.9.011051, PhysRevX.9.021012, guo2019low, 
PhysRevLett.123.050402}, and BECs inside optical resonators~\cite{leonard2017supersolid}.
Additionally, a supersolid-like phase known as the annular stripe phase has also been theoretically
predicted in spin-orbital-angular-momentum-coupled BECs~\cite{chiu2020visible, PhysRevResearch.2.033152, banger2024excitations}.

In SO-coupled BECs, the supersolid stripe phase
exhibits a unique characteristic where atoms condense into a superposition of plane wave states
with finite momenta. This results in density modulations with zero magnetization for identical intraspecies
interactions~\cite{PhysRevLett.105.160403, PhysRevLett.107.150403, PhysRevLett.108.225301, PhysRevLett.110.235302, PhysRevA.86.063621, 
Zheng_2013, PhysRevA.101.043602}.
Other ground-state phases of the SO-coupled BECs are the plane-wave (PW) and zero-momentum (ZM) phases. 
In the PW phase, atoms condense into a single state with finite momenta, which leads to the breaking of
discrete $\mathbb{Z}_2$ symmetry and the emergence of
non-zero magnetizations~\cite{PhysRevA.86.063621}. The PW phase is characterized by a maxon-roton structure
in the elementary excitation spectrum~\cite{PhysRevA.90.063624, PhysRevLett.114.105301}.
Near the transition from the PW to the supersolid ST phase, the roton mode in the excitation spectrum of the PW phase
softens and eventually vanishes at the boundary, marking the transition to the crystalline phase~\cite{PhysRevA.90.063624, PhysRevLett.114.105301}.
The ZM phase, on the other hand, has vanishing momentum and magnetization. 

The zero-temperature phase diagram and elementary excitations of SO-coupled BECs 
have been studied in several works~\cite{PhysRevLett.105.160403, PhysRevLett.107.150403, PhysRevLett.108.225301, PhysRevLett.110.235302, PhysRevA.86.063621, 
Zheng_2013, PhysRevA.93.033648, PhysRevA.93.023615, PhysRevLett.117.125301, PhysRevA.95.033616, PhysRevLett.127.115301, Chen_2022,
PhysRevLett.130.156001, rajat2024collective}. However, research on finite-temperature effects is limited, focusing on pseudo-spinor BECs~\cite{expnature, 
PhysRevA.90.053608, chen2017quantum, PhysRevA.106.023302, PhysRevA.109.033319}. The finite temperature phase transition from the ST to the magnetized PW phase has been 
observed experimentally in a harmonically trapped three-dimensional 
SO-coupled pseudospin-1/2 BEC ~\cite{expnature}. The finite-temperature phase diagram of a homogeneous pseudospin-1/2 SO-coupled BEC has been examined using the Hartree-Fock-
Bogoliubov (HFB) theory with Popov approximation, revealing an increase in the domain of the PW phase at the cost of the ST and ZM phases \cite{chen2017quantum}. 
Within the framework of the HFB theory with Popov approximation,
an examination of the spin-dipole mode of a quasi-one-dimensional harmonically trapped pseudospin-1/2 SO-coupled BEC showed the PW phase crystallizing into 
the supersolid ST phase~\cite{PhysRevA.109.033319}. Previous to these studies on SO-coupled BECs, 
the finite temperature phase diagram of a uniform ferromagnetic spin-1 Bose gas 
without SO coupling has been studied using Hartree-Fock \cite{PhysRevA.85.053611} and HFB theory with Popov approximation~\cite{PhysRevA.84.043645}.

The effects of quantum and thermal fluctuations on the phase diagram of an SO-coupled spin-1 BEC have not been addressed in the literature. In this paper, we undertake this study using the HFB theory with Popov
approximation. The quadratic Zeeman field strength is an additional experimentally controllable parameter that can drive transitions in SO-coupled spin-1 BECs, unlike pseudospin-1/2 BECs. In addition to the single roton-maxon structure in the PW phase, asymmetric/symmetric double roton structures also
emerge in the excitation spectrum of SO-coupled spin-1 BECs \cite{PhysRevA.93.033648, PhysRevA.93.023615}. We measure the roton gap(s) as a function of Raman coupling or quadratic Zeeman field strengths at different temperatures below the critical temperature to delineate the phase boundaries separating the
supersolid phase from the superfluid phases. We also measure the condensate's momentum as a function of Raman coupling strength at different temperatures 
to map out the phase boundary between two superfluid phases. 

This paper has the following structure. In Sec.~\ref{Sec-II}, we discuss the HFB theory with Popov 
approximation for a Raman-induced SO-coupled spin-1 BEC. In Sec.~\ref{Sec-III}, we present and discuss the results.
Finally, we summarize the key findings of this study in Sec.~\ref{Sec-IV}.

\section{Theoretical Model}\label{Sec-II}
We consider a homogeneous three-dimensional (3D) Raman-induced SO-coupled spin-1 
BEC. The grand-canonical Hamiltonian in the second-quantized form for the system is~\cite{NatCommun}
\begin{align}
\mathcal{H}=&\int d\textbf{r}\hat\Psi^{\dagger}\left[\mathcal{L}-\mu+\frac{2\iota \hbar^2 k_R}{m} F_{z}\partial_{x}+ \Omega F_{x}+\left(\epsilon + \frac{2k_R^2\hbar^2}{m}\right)F_{z} ^2\right. \nonumber\\ 
&\left.+ \frac{c_{0}}  {2}  \left(\hat\Psi^{\dagger}\hat\Psi\right) 
+\frac{c_{2}}{2}\left(\hat\Psi^{\dagger}{\bf F}\hat\Psi\right).{\bf F}\right] \hat\Psi, \label{ham}
\end{align}
where $\mathcal{L}=(-\hbar^2 /2m)(\partial_x^2+\partial_y^2+\partial_z^2)$ and $\mu$ is the chemical
potential.
In Eq.~(\ref{ham}), ${\bf F}=(F_{x},F_{y},F_{z})$ are spin-1 Pauli matrices, 
$\hat\Psi=(\hat\Psi_{+1},\hat\Psi_{0},\hat\Psi_{-1})^{\rm T}$ with $\rm T$ denoting the transpose, are the bosonic
annihilating field operators,  $k_R$ is the strength of spin-orbit coupling, $\epsilon$ is the quadratic Zeeman shift, which
can be tuned by detuning the lasers \cite{NatCommun}, $\Omega$ is the Raman coupling strength, which depends on the intensity of
Raman lasers, $c_0$ and $c_2$ are spin-independent and spin-dependent interactions, respectively, which depend on
the $s$-wave scattering lengths \cite{RevModPhys.85.1191}.
A spin-1 BEC with a fixed $c_2/c_0$ admits the ST, the PW, and the ZM phase in $\Omega$-$\epsilon$ plane at absolute zero~\cite{Chen_2022, PhysRevA.93.033648, PhysRevLett.117.125301}. The present work considers a spin-1 BEC with repulsive spin-dependent and spin-independent interactions. 

Considering $1/k_R$, $\hbar^2k_R^2/m$ and $m/\hbar k_R^2$ as the units of length, energy, and time, respectively, we can rewrite Eq.~(\ref{ham}) in
the dimensionless form as
\begin{align}
    \mathcal{H}=\int d\textbf{r}\hat\Psi^{\dagger}\left[\mathcal{L}-\mu+2\iota F_{z}\partial_x+ (\epsilon + 2)F_{z} ^2 + \Omega F_{x} + \frac{c_{0}}  {2} \hat\Psi^{\dagger}\hat\Psi+\frac{c_{2}}{2}\hat\Psi^{\dagger}{\bf F}\hat\Psi.{\bf F} \right] \hat\Psi, \label{h}
\end{align}
where $\mathcal{L}=(\partial_x^2+\partial_y^2+\partial_z^2)/2$.
To address the effects of quantum and thermal fluctuations, we split the field operator as 
$\hat\Psi(\textbf{r},t)= \psi(\textbf{r})+\delta{\hat{\psi}}(\textbf{r},t)$, where $\psi( \textbf{r}) = \langle \hat\Psi(\textbf{r},t) \rangle$
is the condensate wavefunction and $\delta{\hat\psi}(\textbf{r},t)$ is the fluctuation
operator~\cite{PhysRevB.53.9341, PhysRevA.106.013304, PhysRevA.109.033319}. In the PW and ZM phases, the condensate wave function for 
the homogeneous system is of the form
\begin{align*}
\psi( \textbf{r}) = \sqrt{n_{\rm c}}e^{\iota k x}(\alpha_{+1},\alpha_0,\alpha_{-1})^{\rm T},
\end{align*}
where $n_{\rm c}$ is the condensate density, $(\alpha_{+1},\alpha_0,\alpha_{-1})^{\rm T}$
is a normalized spinor with $|\alpha_{+1}|^2 + |\alpha_0|^2 + |\alpha_{-1}|^2 = 1$, and
$k = 2(|\alpha_{+1}|^2 - |\alpha_{-1}|^2)$ is the condensate's momentum, which is non-zero for
the PW and zero for the ZM phase~\cite{PhysRevA.93.023615}.

Using the Bogoliubov transformation, 
$\delta{\hat\psi_i}$ can be expressed in terms of quasi-particle 
creation ($\hat{\beta}_l^{\dagger}$) and annihilation operators ($\hat{\beta}_l$) as
\begin{align}
    \delta{\hat\psi}_{i}(\textbf{r},t)=e^{ \iota k x}\left(\sum_l\left[u_i^l (\textbf{r}) \hat{\beta}_l e^{-\iota \omega_l t}+v_i^{*l}(\textbf{r})\hat{\beta}_l^{\dagger}e^{\iota\omega_l t} \right] \right) \label{fluc},
\end{align}
where $i \in (+1,0,-1)$ represents the component index, 
and $l$ represents the eigenvalue index corresponding to energy $\omega_l$ with $u_i^{l}$ and
$v_i^{l}$ as the quasi-particle wavefunctions. The quasi-particle wavefunctions satisfy the normalization condition
\begin{equation}
\sum_{i}\int d{\bf r}\left[|u_i^l( \textbf{r})|^2 -|v_i^{l}( \textbf{r})|^2\right] = 1,
\end{equation}
and can be expressed in free space as $u_i^l({\bf r}) = u_{i,\textbf{q}}^le^{\iota\textbf{q}.\textbf{r}}/\sqrt{V}$ and $v_i^l({\bf  r}) = v_{i,\textbf{q}}^le^{\iota\textbf{q}.\textbf{r}}/\sqrt{V}$ where $u_{i,\textbf{q}}^l$, $v_{i,\textbf{q}}^l$ are plane-wave amplitudes, $\textbf{q}$ represents the quasi-momentum of the excitation, and $V$ is the total volume of the system.
We consider the Heisenberg equation of motion for the Bose field operator $\hat\Psi(\vec r, t)$, 
\begin{equation}
\iota \frac{\partial}{\partial t} \hat\Psi(\textbf{r}, t)=[\hat\Psi( \textbf{r}, t), \mathcal{H}], 
\label{he}
\end{equation}
and substitute the Hamiltonian (\ref{ham}) and the field operator in terms of the mean field and the fluctuation operator. 
We then apply Wick's theorem, in which the cubic terms in fluctuation operators are decomposed as 
$\delta{\hat\psi}^{\dagger}_{i} \delta{\hat\psi}_{j} \delta{\hat\psi}_{k} \simeq
\left\langle\delta{\hat\psi}^{\dagger}_{i} \delta{\hat\psi}_{j}\right\rangle 
\delta{\hat\psi}_{k}+\left\langle\delta{\hat\psi}^{\dagger}_{i} 
\delta{\hat\psi}_{k}\right\rangle \delta{\hat\psi}_{j}+\langle\delta{\hat\psi}_{j}
\delta{\hat\psi}_{k}\rangle \delta{\hat\psi}^{\dagger}_{i}$, where $\langle\delta{\hat\psi}_{i}\delta{\hat\psi}_{j}\rangle$ is anomalous density term \cite{PhysRevB.53.9341}.
For quadratic terms, we use the mean-field approximation as 
$\delta{\hat\psi}^{\dagger}_{i} \delta{\hat\psi}_{j} \simeq \left\langle\delta{\hat\psi}^{\dagger}_{i} \delta{\hat\psi}_{j}\right\rangle$ 
and $\delta{\hat\psi}_{i} \delta{\hat\psi}_{j} \simeq \left\langle\delta{\hat\psi}_{i} \delta{\hat\psi}_{j}\right\rangle$. Here, we neglect the
anomalous densities in the system to satisfy
the Hugenholtz-Pines theorem \cite{PhysRev.116.489}, known as the Popov approximation.
The average of the three equations of motion gives us the following coupled generalized Gross-Pitaevskii equations 
(GPEs) for the three components of the spinor order parameter  
\begin{subequations}\label{gpe}
\begin{align}
\mu \alpha_{\pm 1}&=  \Bigg[\frac{k^2}{2}+ \epsilon +2 +c_{0}(n+\tilde{n}_{\pm1,\pm1}) \mp 2k +{c_{2}(n_{\pm1}+n_{0}-n_{\mp1}+\tilde{n}_{\pm1,\pm1})}\Bigg]\alpha_{\pm1}\nonumber\\
&+c_{2} n_{\rm c}\alpha_0^2
\alpha_{\mp1}^*+\Bigg[(c_{0}+c_{2})\tilde{n}_{0,\pm1}+2c_{2}\tilde{n}_{\mp1,0} + \frac{\Omega}{\sqrt{2}} \Bigg]\alpha_{0}+(c_{0}-c_{2})\tilde{n}_{\mp1,\pm1}\alpha_{\mp1} ,\\
\mu\alpha_0 &= \Bigg[\frac{k^2}{2}+ c_{0}(n+\tilde{n}_{0,0})+c_{2}(n_{+1}+n_{-1})\Bigg]\alpha_0+\Bigg[(c_{0}+
c_2)\tilde{n}_{+1,0}+2c_2\tilde{n}_{0,-1}+\frac{\Omega}{\sqrt{2}}\Bigg]\alpha_{+1}\nonumber\\
&+\Bigg[(c_{0}+c_2)\tilde{n}_{-1,0}+2c_2\tilde{n}_{0,+1}+\frac{\Omega}{\sqrt{2}}\Bigg] 
\alpha_{-1}+2c_{2}n_{\rm c}\alpha_{+1}\alpha_0^*\alpha_{-1},
\end{align}
\end{subequations}
where $\tilde{n}_{i,j} 
\equiv\langle\delta{\hat\psi}^{\dagger}_{i} \delta{\hat\psi}_{j}\rangle=(1/V)\sum_{l,\textbf{q}}\left\{\left(u_{i,\textbf{q}}^{l *} u_{j,\textbf{q}}^l 
+v_{i,\textbf{q}}^l v_{j,\textbf{q}}^{l *}\right)/(e^{\omega_l/k_{B}T}-1) +v_{i,\textbf{q}}^l v_{j,\textbf{q}}^{l *}\right\}$ with $T$ denoting the temperature.
In Eqs.~(\ref{gpe}a) and (\ref{gpe}b), $n_{i} = n_{{\rm c}i}+\tilde{n}_{i,i}$ is the total density of the $i$th component, where $n_{{\rm c}i} = n_{\rm c}|\alpha_i|^2$ 
is the condensate and $\tilde{n}_{i,i}$ is the non-condensate density, and $n = n_{\rm c}+\tilde{n}$ denotes the total density of the system, where $\tilde{n} = \sum_{i} 
\tilde{n}_{i,i}$ is the total non-condensate density. The non-condensate density $\tilde n$ consists of two contributions, namely thermal depletion $\tilde n_{\rm th} = \sum_i(1/V)\sum_{l,\textbf{q}}\left(|u_{i,\textbf{q}}^{l}|^2 + |v_{i,\textbf{q}}^l|^2\right)/(e^{\omega_l/k_{B}T}-1)$
and quantum depletion $\tilde{n}_{\rm qu} = \sum_i(1/V)\sum_{l,\textbf{q}}|v_{i,\textbf{q}}^l|^2$.
  
For the non-condensate part, we subtract Eqs.~(\ref{gpe}a) and (\ref{gpe}b) from Eq.~(\ref{he}), and then using the Bogoliubov transformation, 
we get the Bogoliubov-de-Gennes (BdG) equations
\begin{equation}\label{bdg}
\begin{pmatrix}
H(q_x,q_y,q_z)+M_1 & \hspace{10pt}n_{\rm c}e^{2\iota k x}M_{2}\\
-n_{\rm c}e^{-2\iota k x}M^*_2  & \hspace{10pt}-H(-q_x,-q_y,-q_z) -M^*_1
\end{pmatrix} 
\begin{pmatrix}
{\mathbf u}_{\mathbf q}^l \\ {\mathbf v}_{\mathbf q}^l 
\end{pmatrix} 
=\omega_l
\begin{pmatrix}
{\mathbf u}_{\mathbf q}^l \\{\mathbf v}_{\mathbf q}^l 
\end{pmatrix},
\end{equation}
where
\begin{align*}
H(q_x,q_y,q_z)&=
\begin{pmatrix}
 h-2(k+q_{x}) + \epsilon +2  \hspace{20pt}& \Omega/\sqrt{2} & 
 0 \\
\Omega/\sqrt{2} & h
&\Omega/\sqrt{2}\\
 0 &  
 \Omega/\sqrt{2} & 
 \hspace{20pt} h+2(k+q_{x})+ \epsilon +2\nonumber
\end{pmatrix},\\
{M}_{1}&=
\begin{pmatrix}
 {c_{0} n_{+1}+c_{2} (2 n_{+1}+n_{0}-n_{-1})} & h_{+1,0} & 
 {(c_{0}-c_{2})(n_c\alpha_{-1}^*\alpha_{+1}+\tilde{n}_{-1,+1})} \\
h_{+1,0}^* & c_{0} n_{0}+{c_{2}(n_{+1}+n_{-1})} &h_{0,-1}\\
 {(c_{0}-c_{2})(n_c\alpha_{+1}^*\alpha_{-1}+\tilde{n}_{+1,-1})} &  
 h_{0,-1}^* & 
 c_{0}n_{-1}+{c_{2}(n_{0}-n_{+1}+2n_{-1})}\nonumber
\end{pmatrix},\\
{M}_{2}&=
\begin{pmatrix}
{(c_{0}+c_{2})\alpha_{+1}^2} & {(c_{0}+c_{2})\alpha_0\alpha_{+1}} & 
{(c_{0}-c_{2})\alpha_{-1}\alpha_{+1}+c_{2}\alpha_0^2} \\
{(c_{0}+c_{2})\alpha_{+1}\alpha_0} &
{c_{0}\alpha_{0}^2+2 c_2 \alpha_{+1} \alpha_{-1}}&
{(c_{0}+c_{2})\alpha_{-1}\alpha_0} \\
{(c_{0}-c_{2})\alpha_{+1}\alpha_{-1}+c_{2}\alpha_0^2} & {(c_{0}+c_{2})\alpha_0\alpha_{-1}} & 
(c_{0}+c_{2})\alpha_{-1}^{2}
\end{pmatrix},
\end{align*}

with $h = (k+q_x)^2/2 +q_y^2/2+q_z^2/2-\mu +c_{0}n, h_{+1,0} = (c_{0}+c_{2})(n_c\alpha_0^*\alpha_{+1}+\tilde{n}_{0,+1})+2 c_2 
(n_c\alpha_{-1}^*\alpha_{0}+\tilde{n}_{-1,0}), h_{0,-1} = (c_{0}+c_{2})(n_c\alpha_{0}\alpha_{-1}^*+\tilde{n}_{-1,0})+2c_2 
(n_c\alpha_{+1}\alpha_{0}^*+\tilde{n}_{0,+1}),{\mathbf u}_{\mathbf q}^l =(u_{+1,\textbf{q}}^l, u_{0,\textbf{q}}^l, u_{-1,\textbf{q}}^l)^{\rm T},~ 
{\mathbf v}_{\mathbf q}^l=(v_{+1,\textbf{q}}^l, v_{0,\textbf{q}}^l, v_{-1,\textbf{q}}^l)^{\rm T}\nonumber$.
We introduce a unitary operator $U$ defined as
\begin{align*}
U = 
\begin{pmatrix}
e^{-\iota k x}I & \hspace{10pt} 0 \\
0  & \hspace{10pt}e^{\iota k x}I
\end{pmatrix}, 
\end{align*}
 where $I$ is a 3$\times$3 identity matrix and transform Eq.~(\ref{bdg}) as
\begin{equation}
U\begin{pmatrix}
H(q_x,q_y,q_z)+M_1 & \hspace{10pt}n_{\rm c}e^{2\iota k x}M_{2}\\
-n_{\rm c}e^{-2\iota k x}M^*_2  & \hspace{10pt}-H(-q_x,-q_y,-q_z) -M^*_1
\end{pmatrix}U^{\dagger}U 
\begin{pmatrix}
{\mathbf u}_{\mathbf q}^l \\ {\mathbf v}_{\mathbf q}^l 
\end{pmatrix} 
=\omega_l U
\begin{pmatrix}
{\mathbf u}_{\mathbf q}^l \\{\mathbf v}_{\mathbf q}^l 
\end{pmatrix}
\end{equation}
to remove the $x$ dependence from the BdG matrix.
The transformed BdG matrix equation now reads as
\begin{equation}\label{BdG_t}
\begin{pmatrix}
H(q_x,q_y,q_z)+M_1 & \hspace{10pt}n_{\rm c}M_{2}\\
-n_{\rm c}M^*_2  & \hspace{10pt}-H(-q_x,-q_y,-q_z) -M^*_1
\end{pmatrix}
\begin{pmatrix}
{\tilde{\mathbf u}}_{\mathbf q}^l \\ {\tilde{\mathbf v}}_{\mathbf q}^l 
\end{pmatrix} 
=\omega_l
\begin{pmatrix}
{\tilde{\mathbf u}}_{\mathbf q}^l \\{\tilde{\mathbf v}}_{\mathbf q}^l 
\end{pmatrix},
\end{equation}
where $(\tilde{\mathbf {u}}_{\mathbf q}^l,\tilde{\mathbf {v}}_{\mathbf q}^l)^{\rm T} = U(\mathbf {u}_{\mathbf q}^l,\mathbf {v}_{\mathbf q}^l)^{\rm T} = (e^{-\iota k x}\mathbf {u}_{\mathbf q}^l,e^{\iota k x}\mathbf {v}_{\mathbf q}^l)^{\rm T}$. 
The term $\tilde{n}_{i,j}$, which remains invariant under this transformation, can be expressed in
terms of transformed Bogoliubov amplitudes as
$\tilde{n}_{i,j} =(1/V)\sum_{l,\textbf{q}}\left\{\left(\tilde{u}_{i,\textbf{q}}^{l *} \tilde{u}_{j,\textbf{q}}^{l}+
\tilde{v}_{i,\textbf{q}}^{l} \tilde{v}_{j,\textbf{q}}^{l *}\right)/(e^{\omega_l/k_{B}T}-1) +\tilde{v}_{i,\textbf{q}}^{l} \tilde{v}_{j,\textbf{q}}^{l *}\right\}.$

\section{Results and Discussions}\label{Sec-III}
We solve Eq.~(\ref{gpe}) and Eq.~(\ref{BdG_t}) numerically using a self-consistent iterative procedure.
In the first step of the self-consistent iterative procedure for a given total density $n = 1$, 
we solve Eq.~(\ref{gpe}) with no spatial dependence using imaginary time propagation.
Here, $\mu$ on the left-hand side of the equation is replaced by $-d/d\tau$ with $\tau$ denoting imaginary time. 
The resultant set of ordinary differential equations can be solved using the forward Euler or Runge-Kutta methods. 
In the second step, transient densities and condensate momentum are provided as input to 
the BdG Eq.~(\ref{BdG_t}). By diagonalizing the $6 \times 6$ BdG matrix, we obtain eigenvalues $\omega_l$ and
eigenvectors $(\mathbf {u}_{\mathbf q}^l \rm,\mathbf {v}_{\mathbf q}^l)^{\rm T}$ for each point on quasi-momentum grid, which are then used to 
construct the non-condensate densities and to renormalize the condensate wavefunctions to ensure that $n_{\rm c} =  n -  \tilde{n}$.
The updated values of $\tilde{n}_{ij}$ and $\alpha_i$ are then fed back into Eq.~(\ref{gpe}) to repeat the
previous two steps. This self-consistent procedure continues until the total non-condensate density achieves a convergence tolerance of $10^{-4}$. To achieve convergence near the phase boundaries, we use the method of successive under-relaxation ~\cite{SIMULA2001396} on the total
thermal density with an under-relaxation parameter $S=0.1$, which implies that $S\tilde{n}^{(p)}+(1-S) \tilde{n}^{(p-1)}$ is updated
as the non-condensate density after $p_{\rm th}$ iteration. We consider a spin-1 BEC with $c_0n = 1.0, c_2n = 0.1$ in a cubical box of volume $L^3$ with periodic boundary 
conditions. To solve the BdG matrix Eq.~(\ref{BdG_t}), we discretize the quasi-momentum space in $N_x \times N_y \times N_z$ grid points with 
discrete $q_\nu$ spanning the  $[(-N_\nu/2+1)2\pi/L, (N_\nu/2)2\pi/L]$ interval with a step size $\Delta q_\nu = 2\pi/L$, 
where $\nu\in(x,y,z)$. This work considers $N_x =  N_y = N_z = 500$.
 
\subsection{Phase diagram in the $T-\Omega$ plane} 
We begin by examining the effect of temperature on the ST-PW and PW-ZM phase boundaries by varying
the Raman coupling strength $\Omega$ while keeping the quadratic Zeeman field fixed at $\epsilon = -1$.

\begin{figure}[!htbp]
\centering
\includegraphics[width=0.9\columnwidth]{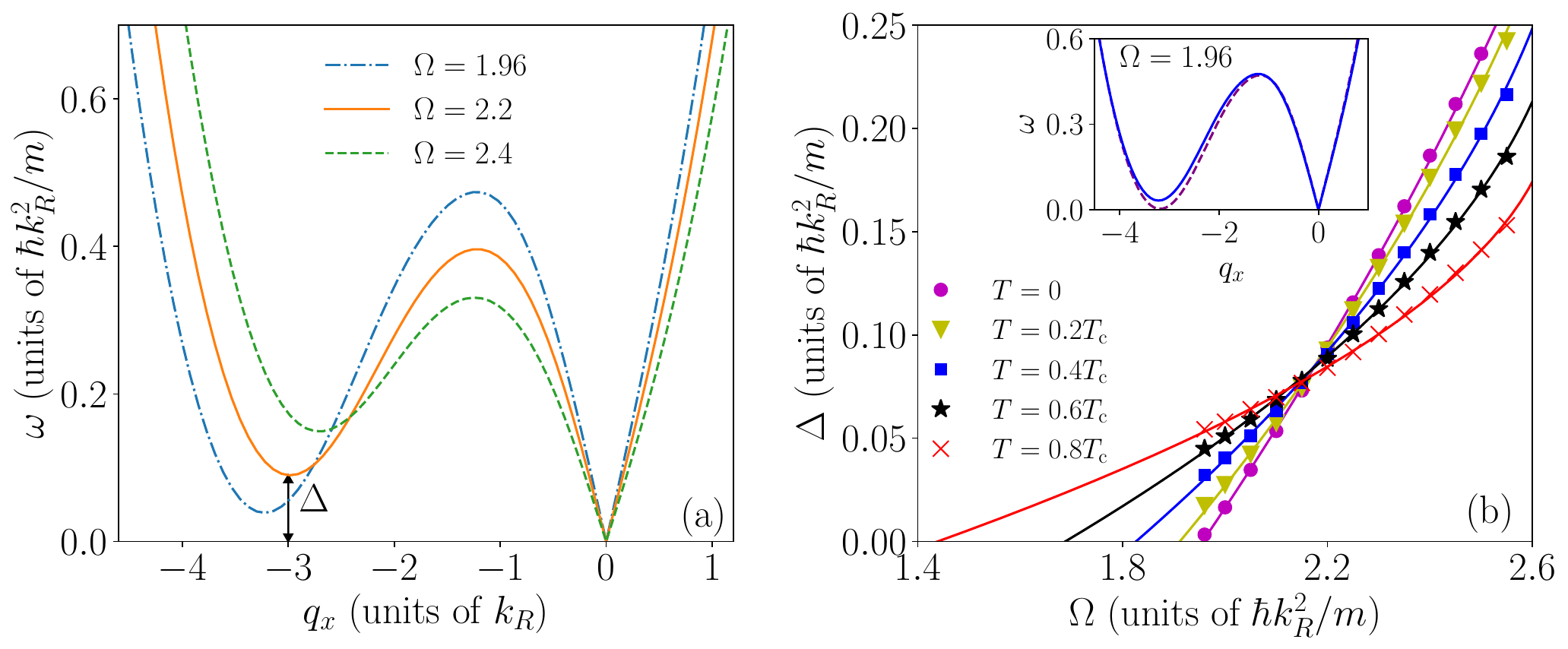}
\caption{(a) The Bogoliubov excitation spectra corresponding to the lowest dispersion band 
at temperature $T = 0.5T_{\rm c}$ with $q_y = q_z = 0 $ for different Raman coupling strengths ($\Omega$).
Here $T_{\rm c} = 2\pi \hbar^2\left[(n/3)/\zeta(3/2)\right]^{2/3}/(mk_B)$ is the transition temperature
for a three-component homogeneous Bose gas. (b) Roton gap $\Delta$ as a function of  $\Omega$ at different temperatures. 
The continuous lines are the quadratic polynomial fits to extract the critical coupling strengths. Inset of (b): the lowest dispersion band for $\Omega = 1.96$ at
$T =0$ (magenta dashed line) and $T=0.4T_{\rm c}$ (blue solid line) showing the opening of the roton gap at
the finite temperature. In (a) and (b), $c_0n = 1$, $c_2n=0.1$, and $\epsilon = -1$.}
\label{roton_PW}
\end{figure}

{\em The ST-PW phase boundary:} We map the ST-PW phase boundary by examining the variation of the roton gap in the dispersion of the PW phase as
a function of the coupling strength at different temperatures. Within the PW phase at a fixed temperature, the roton gap decreases
with a decrease in $\Omega$ and becomes zero at the ST-PW phase boundary ($\Omega_{\rm c_1}$) as a precursor to the crystallization. 
This can be seen in Fig.~\ref{roton_PW}(a), which illustrates the dispersion at different values of $\Omega$ at a fixed temperature 
of $T = 0.5T_{\rm c}$, and Fig.~\ref{roton_PW}(b), which displays the values of the roton gap ($\Delta$) at various temperatures
as $\Omega$ is varied. 
Near the ST-PW phase transition point $\Omega_{\rm c_1}$ in the PW phase, 
the roton gap increases with an increase in temperature. 
Consequently, the roton gap, which closes at $\Omega \approx 1.96$ at absolute zero, opens at a finite temperature [see the inset of Fig~\ref{roton_PW}(b)]. 
This implies a lowering of the critical coupling $\Omega_{\rm c_1}$ as the temperature increases. We extract the roton gap 
from the lowest dispersion band obtained by numerically solving Eqs.~(\ref{gpe}), (\ref{BdG_t}a) and  (\ref{BdG_t}b) as per the self-consistent iterative method up to $\Omega = 1.96$, below which we employ a quadratic polynomial fit to have a measure of $\Omega_{\rm c_1}$.
A similar approach was previously used to calculate the ST-PW phase boundary for a pseudospin-1/2 BEC~\cite{PhysRevLett.114.105301, chen2017quantum}.

\begin{figure}[!htbp]
\centering
\includegraphics[width=0.9\columnwidth]{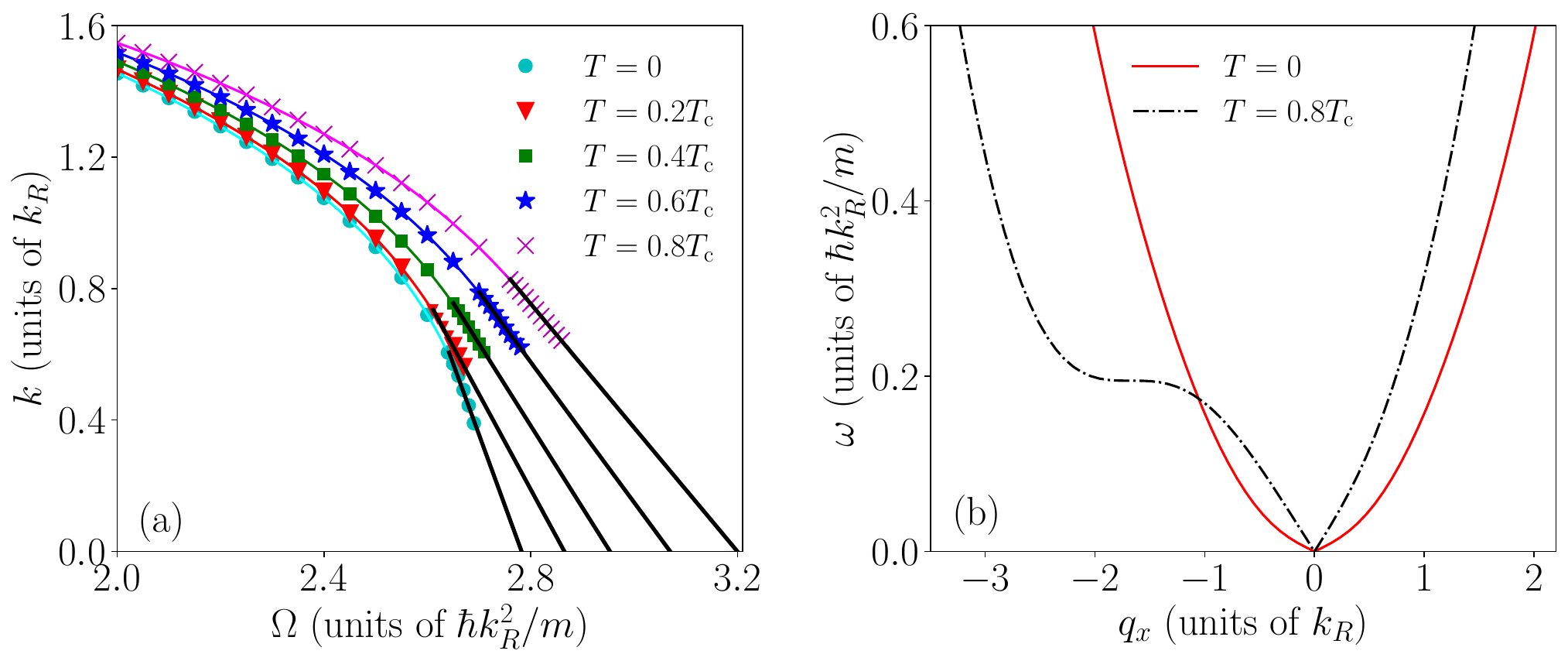}
    \caption{(a) The condensate's momentum in the PW phase as a function of the Raman coupling strength $\Omega$
at different temperatures. The solid black lines denote the linear extrapolation. (b) Lowest dispersion band at temperature $T = 0$ and $T = 0.8T_c$ with $q_y = q_z = 0$ for $\Omega = 2.8$. In (a) and (b), $c_0n = 1$, $c_2n=0.1$, and $\epsilon = -1$.}
    \label{momentum_disp}
\end{figure}

{\em The PW-ZM phase boundary:} The PW phase has a non-zero condensate momentum, which decreases with an increase in $\Omega$ and becomes
zero at the PW-ZM phase boundary ($\Omega_{\rm c_2}$). We, accordingly, use the condensate's momentum $k$ to calculate the the 
PW-ZM phase boundary. In Fig.~\ref{momentum_disp}(a), we show the variation in condensate's momentum in the PW phase with an increase in $\Omega$ at different temperatures. We find that as $\Omega$ approaches the critical point $\Omega_{\rm c_2}$, the self-consistent iterative procedure 
fails to converge due to the enhanced fluctuations. We, therefore, linearly extrapolate $k(\Omega)$ data points to extract the transition point $\Omega_{\rm c_2}$.
The transition point $\Omega_{\rm c_2} \approx 2.78$ at $T=0$, and with an increase in temperature, the transition point shifts 
towards the right. We also examine the lowest dispersion band near the PW-ZM phase transition point $\Omega_{\rm c_2}$ to confirm
the amplification in the PW phase's domain with an increase in temperature. As an example, in Fig.~\ref{momentum_disp}(b), we show the
lowest dispersion band for $\Omega = 2.8$ at $T = 0$ and $0.8T_{\rm c}$. 
The dispersion is symmetric about the $q_x = 0$ at zero temperature in the ZM phase, whereas at $T = 0.8T_{\rm c}$ 
an asymmetric dispersion is illustrative of the PW phase.

\begin{figure}[!htbp]
\centering
\includegraphics[width=0.9\columnwidth]{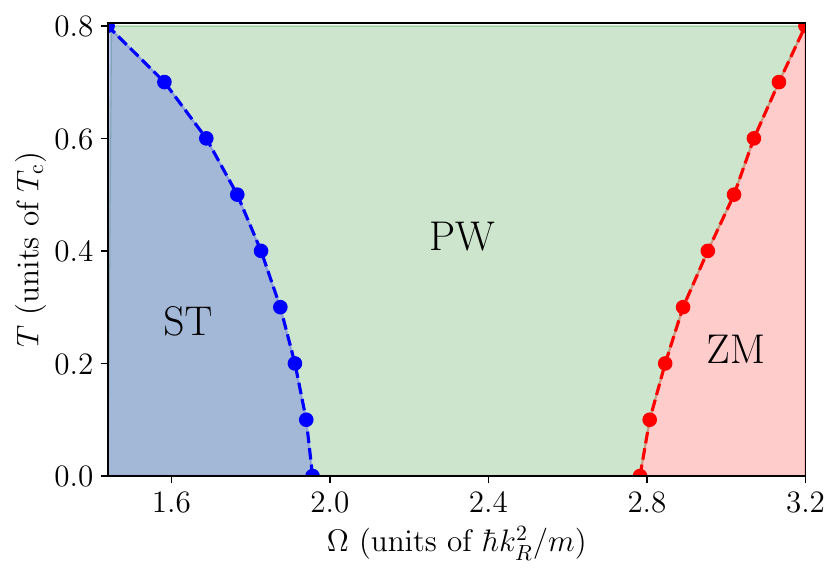}
    \caption{The finite-temperature phase diagram of the Raman-induced SO-coupled spin-1 BEC with $c_0n =1$, $c_2n=0.1$, and $\epsilon = -1$.}
    \label{phase_diagram}
\end{figure}
The complete phase diagram in the $T-\Omega$ plane for $c_0n = 1, c_2 n = 0.1$ and $\epsilon = 1$ is shown in Fig.~\ref{phase_diagram}. 
As the temperature increases, the ST phase melts into the PW phase, as indicated by the ST-PW phase boundary shifting towards the left;
the ZM phase's domain similarly shrinks with the PW-ZM phase boundary shifting towards the right.

\begin{figure}[!htbp]
\centering
\includegraphics[width=0.9\columnwidth]{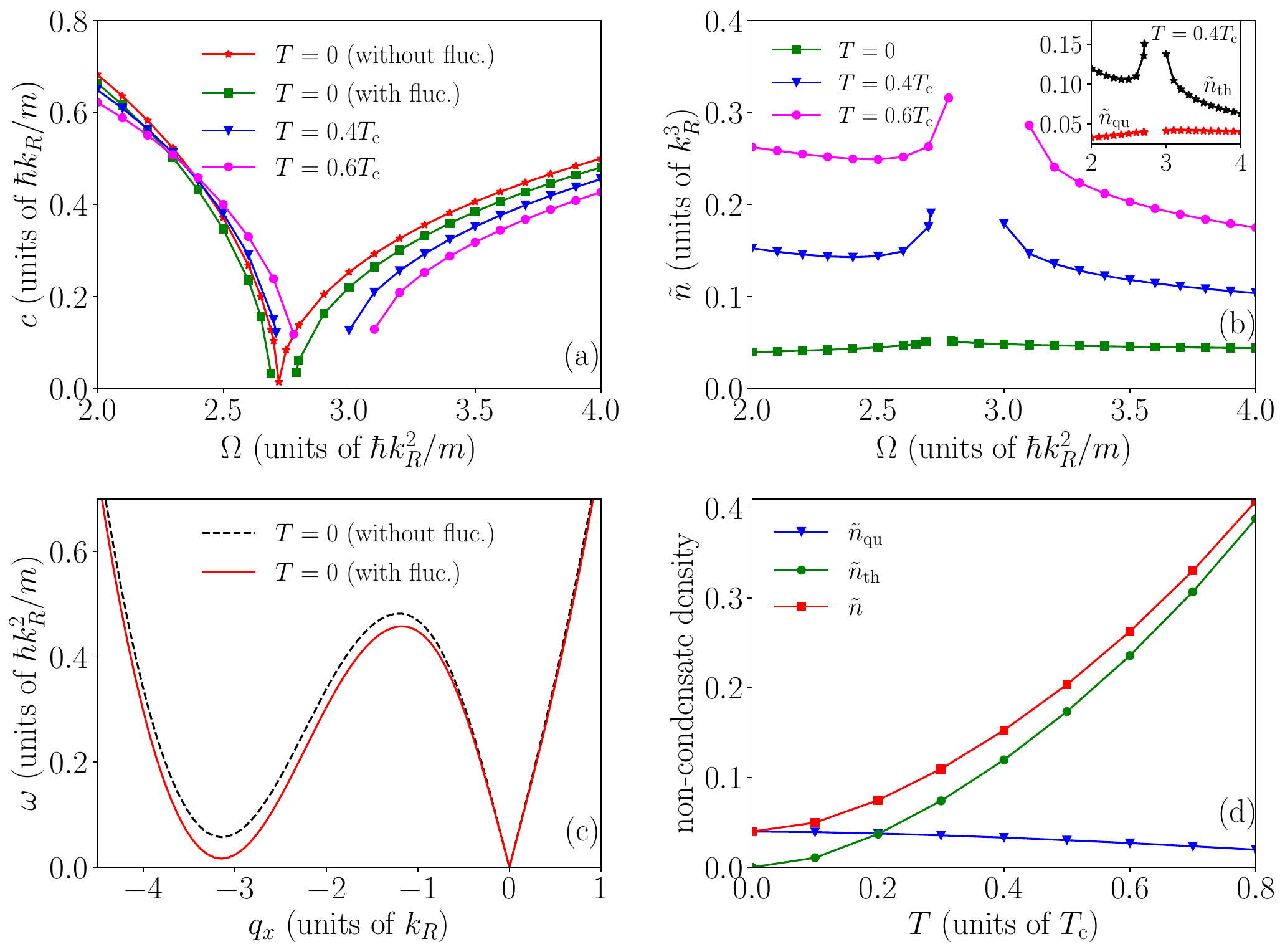}
    \caption{(a) The sound velocities and (b) the non-condensate density as a function of $\Omega$ at different temperatures. Inset of (b) shows
    the thermal ($\tilde{n}_{\rm th}$) and quantum fluctuations' ($\tilde{n}_{\rm qu}$) contributions to $\tilde n = \tilde{n}_{\rm th} + \tilde{n}_{\rm qu}$ as a function of $\Omega$ at $T = 0.4T_{\rm c}$.  
    (c) The lowest dispersion band in the PW phase at zero temperature with and without quantum fluctuation for $\Omega = 2.0$. 
    (d) The contributions to the non-condensate density $\tilde n$ from the thermal ($\tilde{n}_{\rm th}$) and quantum ($\tilde{n}_{\rm qu}$) fluctuations 
       at different temperatures for $\Omega = 2.0$. In (a)-(d) $c_0n = 1$, $c_2n=0.1$, and $\epsilon = -1$.}
    \label{comparison}
\end{figure}

{\em Sound velocities and the condensate depletion:} In the PW phase, an asymmetry in the lowest dispersion band 
about $q_x = 0$, see Fig.~\ref{roton_PW}(a), results in two different sound velocities for $q_x<0$ and $q_x>0$, whereas in 
the ZM phase with a symmetric dispersion the two velocities are equal~\cite{PhysRevLett.114.105301}. In principle, these sound velocities and the condensate depletion 
due to the quantum and thermal fluctuations can be used to estimate the PW-ZM phase boundary. The sound velocities have a minimum, and the condensate depletion has a maximum value at the PW-ZM phase boundary. In Figs.~\ref{comparison}(a) and (b), we present the sound velocities and the non-condensate densities 
at zero and finite temperatures.
For the parameters considered in this work, the sound velocities in the PW phase are approximately equal~\cite{chen2017quantum, PhysRevA.86.063621, Chen_2022, PhysRevLett.114.105301}.

In the ZM phase, as $\Omega$ is decreased, the sound velocity decreases [see Fig.~\ref{comparison}(a)], which leads to an increase 
in the density of the states for the phonon-like excitations. The increased phonon-like excitations at finite temperatures 
lead to an increase in the non-condensate density as $\Omega$ approaches $\Omega_{\rm c_2}$ [see Fig.~\ref{comparison}(b)] \cite{Zheng_2013}.
In the PW phase at finite temperatures, roton and phonon-like excitations contribute to the condensate depletion.
Here, near the ST-PW phase
boundary, the roton gap is small, and the density of states for roton-like excitations is more than the phonon-like
excitations. This results in roton-like excitations primarily contributing to the condensate depletion in this region \cite{Zheng_2013}.
As $\Omega$ increases, the roton-gap increases, making exciting roton-like excitations energetically more costly.
Whereas a decrease in the sound velocity [see Fig.~\ref{comparison}(a)] leads to an increase in the density of states for 
phonon-like excitations, which start contributing more to the depletion. Due to these two competing factors, the non-condensate density 
first decreases and then increases as $\Omega$ is increased \cite{Zheng_2013}. In contrast to the thermal excitations, condensate depletion due to quantum
fluctuations remains almost constant until the self-consistent method breaks down, as is illustrated in the inset of Fig.~\ref{comparison}(b).

{\em Fluctuations and supersolidity:} To examine the role of the quantum fluctuations on supersolidity, 
we calculate the dispersion in the PW phase at $T = 0$ with and without quantum fluctuations.
The quantum fluctuations tend to reduce the roton gap in the PW phase, as shown in Fig.~\ref{comparison}(c), implying a shift
of $\Omega_{\rm c_1}$ to the higher values vis-\`a-vis the prediction of the $T=0$ mean-field model~\cite{Chen_2022}. Fig.~\ref{comparison}(d) shows 
the contribution to non-condensate density from the thermal and quantum fluctuations as the temperature
is increased for a fixed $\Omega$ in the PW phase. The contribution of the quantum fluctuations to the non-condensate
density does not vary much with temperature. In contrast, the thermal fluctuations increase with temperature
as shown in Fig.~\ref{comparison}(d) and shift $\Omega_{\rm c_1}$ to lower values.
The quantum-fluctuation-enhanced supersolidity is in contrast to the thermal fluctuations melting the supersolid ST phase
to the PW phase.

\subsection{The ST-PW phase boundary in $T-\epsilon$ plane}
For a fixed $c_2/c_0$ ratio, the phase transitions in an SO-coupled spin-1 BEC can driven either by varying Raman coupling or 
quadratic Zeeman field strength. Varying the latter to drive the phase transitions is a distinguishing feature of a spin-1 BEC 
vis-\'a-vis pseudospin-1/2 BEC. We now examine the shift in critical Zeeman field $\epsilon_{\rm c_1}$ with temperature for the SO-coupled
spin-1 BEC with $c_0n = 1$, $c_2n =0.1$, and $\Omega = 1.8$, where $\epsilon_{\rm c_1}$ denotes the ST-PW transition point.
Again, we infer the transition point from the closing of the roton gap.
\begin{figure}[!htbp]
\centering
\includegraphics[width=0.9\columnwidth]{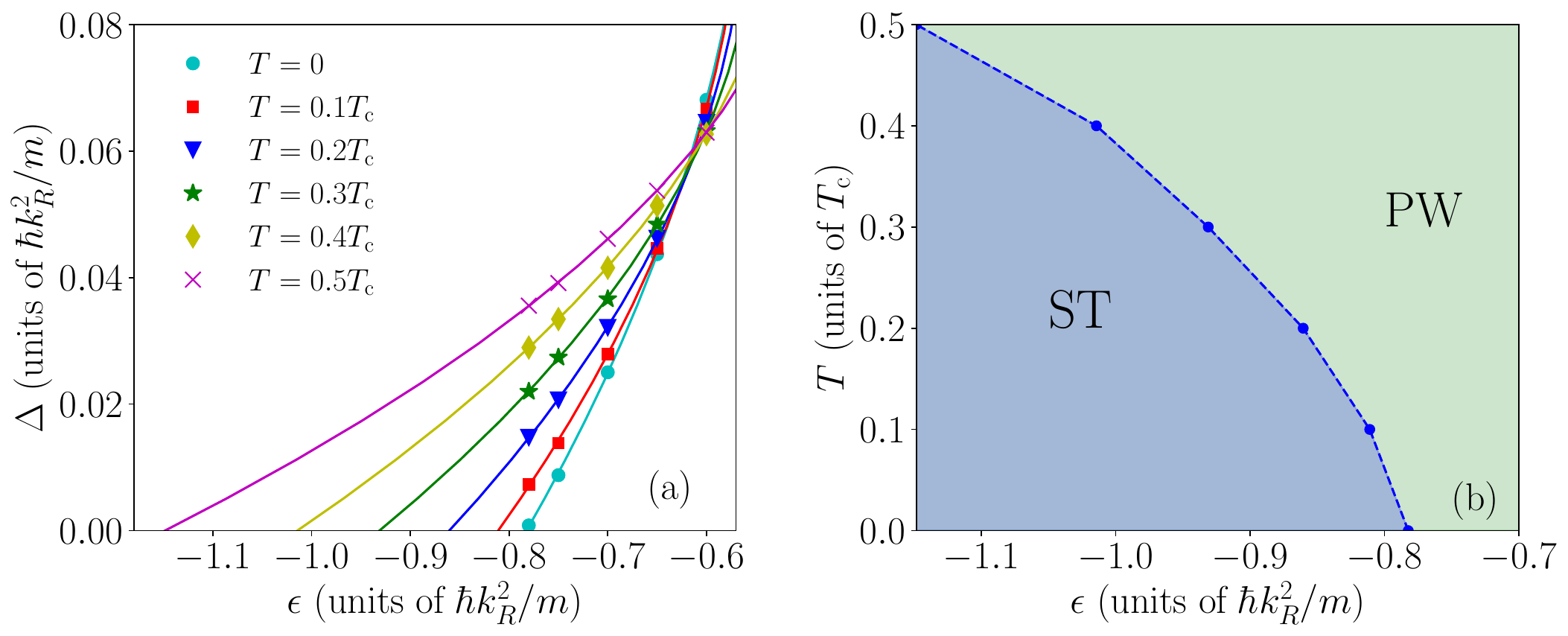}
    \caption{For an SO-coupled BEC with $c_0n = 1$, $c_2n =0.1$, and $\Omega = 1.8$: 
    (a) variation of the roton gap $\Delta$ in the PW phase as function $\epsilon$ at different temperatures, and (b) the ST-PW phase boundary in the $T-\epsilon$ plane.}
    \label{quadratic_zeeman}
\end{figure}
In  Fig.~\ref{quadratic_zeeman}(a), we present the roton gaps at different temperatures as a function of the $\epsilon$ 
in the PW phase. At $T\ne0$ K, we calculated the roton gaps up to $\epsilon=-0.78$,
which is the transition point at $T=0$ K, 
and then used a quadratic polynomial fit to extrapolate and extract $\epsilon_{\rm c_1}$. 
In Fig.~\ref{quadratic_zeeman}(b), we show the ST-PW phase boundary in the $T-\epsilon$ plane.
At zero temperature, the transition point is $\epsilon_{\rm c_1} \approx -0.78$, which shifts towards lower values
as the temperature rises, illustrating the thermal melting of the supersolid ST phase to the PW phase.

\subsection{The ST-ZM phase boundary in $T-\epsilon$ plane}
\begin{figure}[!htbp]
\centering
\includegraphics[width=0.9\columnwidth]{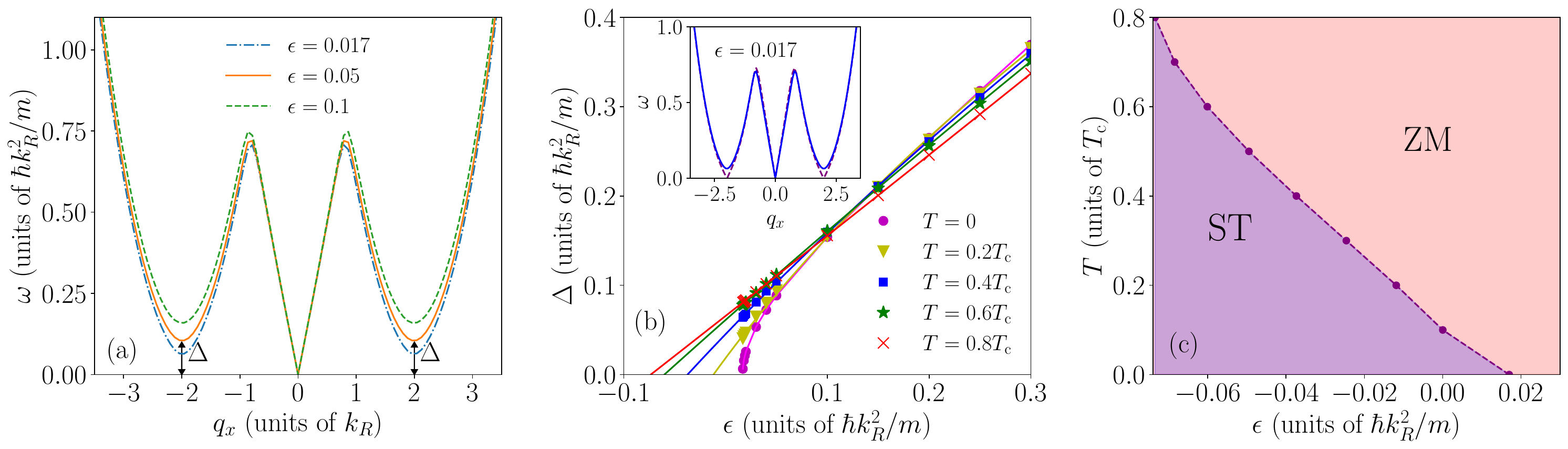}
    \caption{For an SO-coupled BEC with $c_0n = 1$, $c_2n =0.1$, and $\Omega = 0.2$: (a) shows the Bogoliubov excitation spectra
    corresponding to the lowest dispersion band at temperature $T = 0.4T_{\rm{c}}$ with $q_y = q_z = 0 $ for different $\epsilon$; (b) shows the roton gap $\Delta$ as a function of  $\epsilon$ at different temperatures and inset of (b) displays the lowest dispersion band for $\epsilon = 0.017$ at $T =0$ (magenta dashed line) and $T=0.4T_{\rm c}$ (blue solid line), demonstrating the opening of the roton gap at the finite temperature, and (c) depicts the finite temperature phase diagram.}
    \label{st_to_zm}
\end{figure}

For a sufficiently small value of $\Omega$, there is a direct transition from the ZM to the ST phase as one decreases the
quadratic Zeeman field without any intervening PW phase. To examine this, we now consider a spin-1 BEC with $c_0n = 1$, $c_2n =0.1$, and $\Omega = 0.2$ and study
the phase diagram in the $T-\epsilon$ plane. Here, the ZM phase, which occurs above a critical Zeeman field strength, has a distinctive double symmetric roton-maxon structure in its lowest dispersion band \cite{PhysRevA.93.033648, PhysRevA.93.023615}. The symmetric double roton becomes soft as we decrease the quadratic Zeeman field strength, as is shown in Fig.~\ref{st_to_zm}(a) for $T =0.4T_{\rm c}$, and ultimately vanishes at the ST-ZM phase boundary. Fig.~\ref{st_to_zm}(b) shows the roton gap in the ZM phase as a function of the quadratic Zeeman field at different temperatures. We obtain the roton gap from the dispersion for $\epsilon\ge 0.017$ and then use a quadratic polynomial fit to extract
the critical Zeeman field strength. The critical Zeeman field strength decreases with an increase in temperature. 
The finite temperature phase diagram thus obtained is shown in Fig.~\ref{st_to_zm}(c), which indicates the ST phase melting into the ZM phase
as the temperature is raised.  

\section{Conclusions}\label{Sec-IV}
We have implemented the Hartree-Fock-Bogoliubov theory with the Popov approximation to study a Raman-induced spin-orbit coupled spin-1 Bose-Einstein condensate
featuring antiferromagnetic spin-exchange interactions at finite temperatures. 
We calculated the roton gap in the excitation spectrum of the plane-wave phase as a function of Raman coupling ($\Omega$) and the quadratic Zeeman field 
strength ($\epsilon$) at various temperatures ($T$). This allowed us to extract the phase boundary that separates the stripe and plane-wave phases in the $T-\Omega$ and $T-
\epsilon$ planes. Additionally, we tracked the momentum of the condensate in the plane-wave phase as a function of the Raman coupling strength at different temperatures to 
determine the critical coupling strength for the transition to the zero-momentum phase.
At small values of $\Omega$, where a direct transition occurs between the zero-momentum and supersolid stripe phases, the lowest dispersion band in the excitation spectrum of the 
zero-momentum phase displays a characteristic symmetric double roton-maxon structure.
We investigated the disappearance of the symmetric double roton gaps as the quadratic Zeeman field strength is lowered, which allowed us to identify the transition point.
We observed that the supersolid stripe phase melts into the superfluid plane-wave or zero-momentum phase as the temperature increases. Furthermore, we demonstrated that 
quantum and thermal fluctuations affect the boundary separating the supersolid stripe from the superfluid phases in different ways. Specifically, quantum fluctuations tend 
to amplify supersolidity, while thermal fluctuations have the opposite effect. 

\section*{Acknowledgments}
A.R. acknowledges the support of the Science and Engineering Research Board
(SERB), Department of Science and Technology, Government of India, under the project SRG/2022/000057
and IIT Mandi seed-grant funds under the project
IITM/SG/AR/87. A.R. acknowledges the National Supercomputing Mission (NSM) for providing computing resources of PARAM Himalaya at IIT Mandi,
which is implemented by C-DAC and supported by the Ministry of Electronics and Information Technology (MeitY) and Department of Science
and Technology (DST), Government of India. S.G. acknowledges support from the Science and Engineering Research Board, Department
of Science and Technology, Government of India through Project No. CRG/2021/002597.

\bibliographystyle{iopart-num} 
\bibliography{spin1}  

\providecommand{\newblock}{}
\begin{thebibliography}{10}
\expandafter\ifx\csname url\endcsname\relax
  \def\url#1{{\tt #1}}\fi
\expandafter\ifx\csname urlprefix\endcsname\relax\def\urlprefix{URL }\fi
\providecommand{\eprint}[2][]{\url{#2}}

\bibitem{galitski2013spin}
Galitski V and Spielman I~B 2013 Spin-orbit coupling in quantum gases {\em Nature\/} {\bf \href{https://doi.org/10.1038/nature11841}{494}} \href{https://doi.org/10.1038/nature11841}{49--54}

\bibitem{Goldman_2014}
Goldman N, Juzeliūnas G, Öhberg P and Spielman I~B 2014 Light-induced gauge fields for ultracold atoms {\em Reports on Progress in Physics\/} {\bf \href{https://dx.doi.org/10.1088/0034-4885/77/12/126401}{77}} \href{https://dx.doi.org/10.1088/0034--4885/77/12/126401}{126401}

\bibitem{sorealize}
Lin Y, Jiménez-García K and Spielman I~B 2011 Spin–orbit-coupled Bose–Einstein condensates {\em Nature\/} {\bf \href{https://doi.org/10.1038/nature09887}{471}} \href{https://doi.org/10.1038/nature09887}{83–86}

\bibitem{NatCommun}
Campbell D, Price R, Putra A, Vald{\'e}s-Curiel A, Trypogeorgos D and Spielman I~B 2016 Magnetic phases of spin-1 spin–orbit-coupled Bose gases {\em Nat. Commun.\/} {\bf \href{https://doi.org/10.1038/ncomms10897}{7}} \href{https://doi.org/10.1038/ncomms10897}{10897}

\bibitem{synthesis_so}
Luo X, Wu L, Chen J and et~al 2016 Tunable atomic spin-orbit coupling synthesized with a modulating gradient magnetic field {\em Sci. Rep.\/} {\bf \href{https://doi.org/10.1038/srep18983}{6}} \href{https://doi.org/10.1038/srep18983}{18983}

\bibitem{recati2023supersolidity}
Recati A and Stringari S 2023 Supersolidity in ultracold dipolar gases {\em Nature Reviews Physics\/} {\bf \href{https://doi.org/10.1038/s42254-023-00648-2}{5}} \href{https://doi.org/10.1038/s42254--023--00648--2}{735--743}

\bibitem{li2017stripe}
Li J~R, Lee J, Huang W, Burchesky S, Shteynas B, Top F~{\c{C}}, Jamison A~O and Ketterle W 2017 A stripe phase with supersolid properties in spin--orbit-coupled Bose-Einstein condensates {\em Nature\/} {\bf \href{https://doi.org/10.1038/nature21431}{543}} \href{https://doi.org/10.1038/nature21431}{91--94}

\bibitem{PhysRevLett.124.053605}
Putra A, Salces-C\'arcoba F, Yue Y, Sugawa S and Spielman I~B 2020 Spatial Coherence of Spin-Orbit-Coupled Bose Gases {\em Phys. Rev. Lett.\/} {\bf \href{https://link.aps.org/doi/10.1103/PhysRevLett.124.053605}{124}} \href{https://link.aps.org/doi/10.1103/PhysRevLett.124.053605}{053605}

\bibitem{chisholm2024probing}
Chisholm C, Hirthe S, Makhalov V, Ramos R, Vatr{\'e} R, Cabedo J, Celi A and Tarruell L 2024 Probing supersolidity through excitations in a spin-orbit-coupled Bose-Einstein condensate {\em \href{https://doi.org/10.48550/arXiv.2412.13861}{arXiv:2412.13861}\/}

\bibitem{10.21468/SciPostPhys.11.5.092}
Martone G~I and Stringari S 2021 Supersolid phase of a spin-orbit-coupled Bose-Einstein condensate: A perturbation approach {\em SciPost Phys.\/} {\bf \href{https://scipost.org/10.21468/SciPostPhys.11.5.092}{11}} \href{https://scipost.org/10.21468/SciPostPhys.11.5.092}{092}

\bibitem{PhysRevLett.122.130405}
Tanzi L, Lucioni E, Fam\`a F, Catani J, Fioretti A, Gabbanini C, Bisset R~N, Santos L and Modugno G 2019 Observation of a Dipolar Quantum Gas with Metastable Supersolid Properties {\em Phys. Rev. Lett.\/} {\bf \href{https://link.aps.org/doi/10.1103/PhysRevLett.122.130405}{122}} \href{https://link.aps.org/doi/10.1103/PhysRevLett.122.130405}{130405}

\bibitem{PhysRevX.9.011051}
B\"ottcher F, Schmidt J~N, Wenzel M, Hertkorn J, Guo M, Langen T and Pfau T 2019 Transient Supersolid Properties in an Array of Dipolar Quantum Droplets {\em Phys. Rev. X\/} {\bf \href{https://link.aps.org/doi/10.1103/PhysRevX.9.011051}{9}} \href{https://link.aps.org/doi/10.1103/PhysRevX.9.011051}{011051}

\bibitem{PhysRevX.9.021012}
Chomaz L, Petter D, Ilzh\"ofer P, Natale G, Trautmann A, Politi C, Durastante G, van Bijnen R~M~W, Patscheider A, Sohmen M, Mark M~J and Ferlaino F 2019 Long-Lived and Transient Supersolid Behaviors in Dipolar Quantum Gases {\em Phys. Rev. X\/} {\bf \href{https://link.aps.org/doi/10.1103/PhysRevX.9.021012}{9}} \href{https://link.aps.org/doi/10.1103/PhysRevX.9.021012}{021012}

\bibitem{guo2019low}
Guo M, B{\"o}ttcher F, Hertkorn J, Schmidt J~N, Wenzel M, B{\"u}chler H~P, Langen T and Pfau T 2019 The low-energy Goldstone mode in a trapped dipolar supersolid {\em Nature\/} {\bf \href{https://doi.org/10.1038/s41586-019-1569-5}{574}} \href{https://doi.org/10.1038/s41586--019--1569--5}{386--389}

\bibitem{PhysRevLett.123.050402}
Natale G, van Bijnen R~M~W, Patscheider A, Petter D, Mark M~J, Chomaz L and Ferlaino F 2019 Excitation Spectrum of a Trapped Dipolar Supersolid and Its Experimental Evidence {\em Phys. Rev. Lett.\/} {\bf \href{https://link.aps.org/doi/10.1103/PhysRevLett.123.050402}{123}} \href{https://link.aps.org/doi/10.1103/PhysRevLett.123.050402}{050402}

\bibitem{leonard2017supersolid}
L{\'e}onard J, Morales A, Zupancic P, Esslinger T and Donner T 2017 Supersolid formation in a quantum gas breaking a continuous translational symmetry {\em Nature\/} {\bf \href{https://doi.org/10.1038/nature21067}{543}} \href{https://doi.org/10.1038/nature21067}{87--90}

\bibitem{chiu2020visible}
Chiu N, Kawaguchi Y, Yip S and Lin Y 2020 Visible stripe phases in spin--orbital-angular-momentum coupled Bose--Einstein condensates {\em New J. Phys.\/} {\bf \href{https://dx.doi.org/10.1088/1367-2630/abac3c}{22}} \href{https://dx.doi.org/10.1088/1367--2630/abac3c}{093017}

\bibitem{PhysRevResearch.2.033152}
Chen X~L, Peng S~G, Zou P, Liu X~J and Hu H 2020 Angular stripe phase in spin-orbital-angular-momentum coupled Bose condensates {\em Phys. Rev. Res.\/} {\bf \href{https://link.aps.org/doi/10.1103/PhysRevResearch.2.033152}{2}} \href{https://link.aps.org/doi/10.1103/PhysRevResearch.2.033152}{033152}

\bibitem{banger2024excitations}
Banger P, Rajat and Gautam S 2024 Excitations of a supersolid annular stripe phase in a spin-orbital-angular-momentum-coupled spin-1 Bose-Einstein condensate {\em \href{https://doi.org/10.48550/arXiv.2411.17586}{arXiv:2411.17586}\/}

\bibitem{PhysRevLett.105.160403}
Wang C, Gao C, Jian C~M and Zhai H 2010 Spin-Orbit Coupled Spinor Bose-Einstein Condensates {\em Phys. Rev. Lett.\/} {\bf \href{https://link.aps.org/doi/10.1103/PhysRevLett.105.160403}{105}} \href{https://link.aps.org/doi/10.1103/PhysRevLett.105.160403}{160403}

\bibitem{PhysRevLett.107.150403}
Ho T~L and Zhang S 2011 Bose-Einstein Condensates with Spin-Orbit Interaction {\em Phys. Rev. Lett.\/} {\bf \href{https://link.aps.org/doi/10.1103/PhysRevLett.107.150403}{107}} \href{https://link.aps.org/doi/10.1103/PhysRevLett.107.150403}{150403}

\bibitem{PhysRevLett.108.225301}
Li Y, Pitaevskii L~P and Stringari S 2012 Quantum Tricriticality and Phase Transitions in Spin-Orbit Coupled Bose-Einstein Condensates {\em Phys. Rev. Lett.\/} {\bf \href{https://link.aps.org/doi/10.1103/PhysRevLett.108.225301}{108}} \href{https://link.aps.org/doi/10.1103/PhysRevLett.108.225301}{225301}

\bibitem{PhysRevLett.110.235302}
Li Y, Martone G~I, Pitaevskii L~P and Stringari S 2013 Superstripes and the Excitation Spectrum of a Spin-Orbit-Coupled Bose-Einstein Condensate {\em Phys. Rev. Lett.\/} {\bf \href{https://link.aps.org/doi/10.1103/PhysRevLett.110.235302}{110}} \href{https://link.aps.org/doi/10.1103/PhysRevLett.110.235302}{235302}

\bibitem{PhysRevA.86.063621}
Martone G~I, Li Y, Pitaevskii L~P and Stringari S 2012 Anisotropic dynamics of a spin-orbit-coupled Bose-Einstein condensate {\em Phys. Rev. A\/} {\bf \href{https://link.aps.org/doi/10.1103/PhysRevA.86.063621}{86}} \href{https://link.aps.org/doi/10.1103/PhysRevA.86.063621}{063621}

\bibitem{Zheng_2013}
Zheng W, Yu Z~Q, Cui X and Zhai H 2013 Properties of Bose gases with the Raman-induced spin–orbit coupling {\em J. Phys. B\/} {\bf \href{https://dx.doi.org/10.1088/0953-4075/46/13/134007}{46}} \href{https://dx.doi.org/10.1088/0953--4075/46/13/134007}{134007}

\bibitem{PhysRevA.101.043602}
S\'anchez-Baena J, Boronat J and Mazzanti F 2020 Supersolid stripes enhanced by correlations in a Raman spin-orbit-coupled system {\em Phys. Rev. A\/} {\bf \href{https://link.aps.org/doi/10.1103/PhysRevA.101.043602}{101}} \href{https://link.aps.org/doi/10.1103/PhysRevA.101.043602}{043602}

\bibitem{PhysRevA.90.063624}
Khamehchi M~A, Zhang Y, Hamner C, Busch T and Engels P 2014 Measurement of collective excitations in a spin-orbit-coupled Bose-Einstein condensate {\em Phys. Rev. A\/} {\bf \href{https://link.aps.org/doi/10.1103/PhysRevA.90.063624}{90}} \href{https://link.aps.org/doi/10.1103/PhysRevA.90.063624}{063624}

\bibitem{PhysRevLett.114.105301}
Ji S~C, Zhang L, Xu X~T, Wu Z, Deng Y, Chen S and Pan J~W 2015 Softening of Roton and Phonon Modes in a Bose-Einstein Condensate with Spin-Orbit Coupling {\em Phys. Rev. Lett.\/} {\bf \href{https://link.aps.org/doi/10.1103/PhysRevLett.114.105301}{114}} \href{https://link.aps.org/doi/10.1103/PhysRevLett.114.105301}{105301}

\bibitem{PhysRevA.93.033648}
Yu Z~Q 2016 Phase transitions and elementary excitations in spin-1 Bose gases with Raman-induced spin-orbit coupling {\em Phys. Rev. A\/} {\bf \href{https://link.aps.org/doi/10.1103/PhysRevA.93.033648}{93}} \href{https://link.aps.org/doi/10.1103/PhysRevA.93.033648}{033648}

\bibitem{PhysRevA.93.023615}
Sun K, Qu C, Xu Y, Zhang Y and Zhang C 2016 Interacting spin-orbit-coupled spin-1 Bose-Einstein condensates {\em Phys. Rev. A\/} {\bf \href{https://link.aps.org/doi/10.1103/PhysRevA.93.023615}{93}} \href{https://link.aps.org/doi/10.1103/PhysRevA.93.023615}{023615}

\bibitem{PhysRevLett.117.125301}
Martone G~I, Pepe F~V, Facchi P, Pascazio S and Stringari S 2016 Tricriticalities and Quantum Phases in Spin-Orbit-Coupled Spin-1 Bose Gases {\em Phys. Rev. Lett.\/} {\bf \href{https://link.aps.org/doi/10.1103/PhysRevLett.117.125301}{117}} \href{https://link.aps.org/doi/10.1103/PhysRevLett.117.125301}{125301}

\bibitem{PhysRevA.95.033616}
Chen L, Pu H, Yu Z~Q and Zhang Y 2017 Collective excitation of a trapped Bose-Einstein condensate with spin-orbit coupling {\em Phys. Rev. A\/} {\bf \href{https://link.aps.org/doi/10.1103/PhysRevA.95.033616}{95}} \href{https://link.aps.org/doi/10.1103/PhysRevA.95.033616}{033616}

\bibitem{PhysRevLett.127.115301}
Geier K~T, Martone G~I, Hauke P and Stringari S 2021 Exciting the Goldstone Modes of a Supersolid Spin-Orbit-Coupled Bose Gas {\em Phys. Rev. Lett.\/} {\bf \href{https://link.aps.org/doi/10.1103/PhysRevLett.127.115301}{127}} \href{https://link.aps.org/doi/10.1103/PhysRevLett.127.115301}{115301}

\bibitem{Chen_2022}
Chen Y, Lyu H, Xu Y and Zhang Y 2022 Elementary excitations in a spin–orbit-coupled spin-1 Bose–Einstein condensate {\em New J. Phys.\/} {\bf \href{https://dx.doi.org/10.1088/1367-2630/ac7fb1}{24}} \href{https://dx.doi.org/10.1088/1367--2630/ac7fb1}{073041}

\bibitem{PhysRevLett.130.156001}
Geier K~T, Martone G~I, Hauke P, Ketterle W and Stringari S 2023 Dynamics of Stripe Patterns in Supersolid Spin-Orbit-Coupled Bose Gases {\em Phys. Rev. Lett.\/} {\bf \href{https://link.aps.org/doi/10.1103/PhysRevLett.130.156001}{130}} \href{https://link.aps.org/doi/10.1103/PhysRevLett.130.156001}{156001}

\bibitem{rajat2024collective}
Rajat, Banger P and Gautam S 2024 Collective excitations and universal coarsening dynamics of a spin-orbit-coupled spin-1 Bose-Einstein condensates {\em \href{https://doi.org/10.48550/arXiv.2410.22178}{arXiv:2410.22178}\/}

\bibitem{expnature}
Ji S~C, Zhang J~Y, Zhang L, Du Z~D, Zheng W, Deng Y~J, Zhai H, Chen S and Pan J~W 2014 Experimental determination of the finite-temperature phase diagram of a spin–orbit coupled Bose gas {\em Nature Phys.\/} {\bf \href{https://doi.org/10.1038/nphys2905 }{10}} \href{https://doi.org/10.1038/nphys2905 }{314–320}

\bibitem{PhysRevA.90.053608}
Yu Z~Q 2014 Equation of state and phase transition in spin-orbit-coupled Bose gases at finite temperature: A perturbation approach {\em Phys. Rev. A\/} {\bf \href{https://link.aps.org/doi/10.1103/PhysRevA.90.053608}{90}} \href{https://link.aps.org/doi/10.1103/PhysRevA.90.053608}{053608}

\bibitem{chen2017quantum}
Chen X~L, Liu X~J and Hu H 2017 Quantum and thermal fluctuations in a Raman spin-orbit-coupled Bose gas {\em Physical Review A\/} {\bf \href{https://link.aps.org/doi/10.1103/PhysRevA.96.013625}{96}} \href{https://link.aps.org/doi/10.1103/PhysRevA.96.013625}{013625}

\bibitem{PhysRevA.106.023302}
Chen X~L, Liu X~J and Hu H 2022 Superfluidity of a Raman spin-orbit-coupled Bose gas at finite temperature {\em Phys. Rev. A\/} {\bf \href{https://link.aps.org/doi/10.1103/PhysRevA.106.023302}{106}} \href{https://link.aps.org/doi/10.1103/PhysRevA.106.023302}{023302}

\bibitem{PhysRevA.109.033319}
Rajat, Ritu, Roy A and Gautam S 2024 Temperature-induced supersolidity in spin-orbit-coupled Bose gases {\em Phys. Rev. A\/} {\bf \href{https://link.aps.org/doi/10.1103/PhysRevA.109.033319}{109}} \href{https://link.aps.org/doi/10.1103/PhysRevA.109.033319}{033319}

\bibitem{PhysRevA.85.053611}
Kawaguchi Y, Phuc N~T and Blakie P~B 2012 Finite-temperature phase diagram of a spin-1 Bose gas {\em Phys. Rev. A\/} {\bf \href{https://link.aps.org/doi/10.1103/PhysRevA.85.053611}{85}} \href{https://link.aps.org/doi/10.1103/PhysRevA.85.053611}{053611}

\bibitem{PhysRevA.84.043645}
Phuc N~T, Kawaguchi Y and Ueda M 2011 Effects of thermal and quantum fluctuations on the phase diagram of a spin-1 ${}^{87}$Rb Bose-Einstein condensate {\em Phys. Rev. A\/} {\bf \href{https://link.aps.org/doi/10.1103/PhysRevA.84.043645}{84}} \href{https://link.aps.org/doi/10.1103/PhysRevA.84.043645}{043645}

\bibitem{RevModPhys.85.1191}
Stamper-Kurn D~M and Ueda M 2013 Spinor Bose gases: Symmetries, magnetism, and quantum dynamics {\em Rev. Mod. Phys.\/} {\bf \href{https://link.aps.org/doi/10.1103/RevModPhys.85.1191}{85}} \href{https://link.aps.org/doi/10.1103/RevModPhys.85.1191}{1191--1244}

\bibitem{PhysRevB.53.9341}
Griffin A 1996 Conserving and gapless approximations for an inhomogeneous Bose gas at finite temperatures {\em Phys. Rev. B\/} {\bf \href{https://link.aps.org/doi/10.1103/PhysRevB.53.9341}{53}} \href{https://link.aps.org/doi/10.1103/PhysRevB.53.9341}{9341--9347}

\bibitem{PhysRevA.106.013304}
Rajat, Roy A and Gautam S 2022 Collective excitations in cigar-shaped spin-orbit-coupled spin-1 Bose-Einstein condensates {\em Phys. Rev. A\/} {\bf \href{https://link.aps.org/doi/10.1103/PhysRevA.106.013304}{106}} \href{https://link.aps.org/doi/10.1103/PhysRevA.106.013304}{013304}

\bibitem{PhysRev.116.489}
Hugenholtz N~M and Pines D 1959 Ground-State Energy and Excitation Spectrum of a System of Interacting Bosons {\em Phys. Rev.\/} {\bf \href{https://link.aps.org/doi/10.1103/PhysRev.116.489}{116}} \href{https://link.aps.org/doi/10.1103/PhysRev.116.489}{489--506}

\bibitem{SIMULA2001396}
Simula T, Virtanen S and Salomaa M 2001 Quantized circulation in dilute Bose–Einstein condensates {\em Computer Physics Communications\/} {\bf \href{https://www.sciencedirect.com/science/article/pii/S0010465501003691}{142}} \href{https://www.sciencedirect.com/science/article/pii/S0010465501003691}{396--400}

\end{thebibliography}
\end{document}